\documentclass[11pt]{article}
\usepackage
{geometry}
\usepackage{amsmath,amssymb}
\usepackage[utf8x]{inputenc}
\usepackage{nameref}  
\usepackage[right]{lineno}
\usepackage{microtype}
\DisableLigatures[f]{encoding = *, family = * }
\usepackage[table]{xcolor}
\usepackage{array}
\newlength\savedwidth

\raggedright
\usepackage[aboveskip=1pt,labelfont=bf,labelsep=period,justification=raggedright,singlelinecheck=off]{caption}

\makeatletter
\renewcommand{\@biblabel}[1]{\quad#1.}
\makeatother
\usepackage{lastpage,fancyhdr,graphicx}
\usepackage{epstopdf}
\usepackage{sfsbib}
\fancyheadoffset[L]{2.25in}
\fancyfootoffset[L]{2.25in}
\usepackage{replterms}
\newcommand{\etal}{\textit{et al. }}
\newcommand{\Figures}{y}
\usepackage{graphicx}
\begin{document}
\baselineskip15pt

\title{Replicability: Terminology, Measuring Success, and Strategy}
\author{
Werner A. Stahel
\\
\bigskip
Seminar for Statistics, ETH, Zurich, Switzerland;
stahel@stat.math.ethz.ch
}
\maketitle

\section*{Abstract}
Empirical science needs to be based on facts and claims that can be
reproduced.
This calls for replicating the studies that proclaim the claims,
but practice in most fields still fails to implement this idea.
When such studies emerged in the past decade, the results were generally
disappointing.
There have been an overwhelming number of papers addressing the
``reproducibility crisis'' in the last 20 years.
Nevertheless, terminology is not yet settled, and
there is no consensus about when a replication should be called
successful.
This paper intends to clarify such issues.

A fundamental problem in empirical science is that usual claims only state
that effects are non-zero, and such statements are scientifically void.
An effect must have a \emph{relevant} size to become a reasonable item
of knowledge. Therefore, estimation of an effect, with an indication of
precision, forms a substantial scientific task, whereas testing it against
zero does not. 
A relevant effect is one that is shown to exceed a relevance threshold.
This paradigm has implications for the judgement on replication success. 

A further issue is the unavoidable variability between studies,
called heterogeneity in meta-analysis.
Therefore, it is of little value, again, to test for zero difference
between an original effect and its replication, but exceedance of a
corresponding relevance threshold should be tested.
In order to estimate the degree of heterogeneity,
more than one replication is needed, and an appropriate indication of the
precision of an estimated effect requires such an estimate.

These insights, which are discussed in the paper, show
the complexity of obtaining solid scientific results,
implying the need for a strategy to make replication happen.
\\

\textbf{KEYWORDS} \qquad
reproducibility; replicability; confirmability.\\

\textbf{Note.}
This paper has been finalized in Feb. 2025 and is now, Aug. 2025, stored in arXiv.

\pagebreak
\section{Introduction}

Science is supposed to be based on facts that are reproducible---or, as we
will call it below, \Term{confirmable}:
If the same phenomenon is studied again with the same methods, the
essential results or conclusions should be the same.
Such validation encompasses all kinds and steps (Section \ref{sec:terms}).
In empirical science, claims are based on data subject to random variation.
Performing a whole study again is called a \Term{replication},
and such projects have been rare in many fields.
(For psychology, see \cite{MakMPH12}.)


The literature on replicability focusses on studies assessing
a quantitative or qualitative ``effect,''
and this will also be the topic here.

\subsection{The replication crisis}
In the last ten years, there have been ample studies
in diverse fields of science 
that found replications to fail in too many instances,
and this experience has lead to such a vast literature that we limit
ourselves to citing the book by the
\citeasnoun{NatAcad19} for a general introduction.

A wealth of papers have argued and speculated about the
reasons for the widespread failure of replicability.
Considering empirical research,
this author is convinced that the dominant cause
in many fields is the so-called \Term{selective reporting bias}
(or, more simply, selection bias, \cite[p.1328]{StaTCD18})
at various levels, such as:
\begin{itemize}
\item 
  Data dredging, P hacking and 
  HARK-ing, that is, Hypothesizing After the Results are Known
  (Acad.\ Med.\ Sci., 2015\nocite{AMedSci15}, and others):
  searching for patterns without a pre-conceived question or hypothesis,
  picking the most salient ones,
  and then issuing a statistical hypothesis for the most salient
  one(s), followed by formal statistical testing;
  
\item
  \Term{garden of forking paths}  (\cite{GelAL14})
  or \Term{researcher degrees of freedom} (\cite{SimJNS11})
  selecting, among different possibilities of analyzing the data,
  the one that gives the largest effect;
\item
  \Term{reporting bias} (\cite{DwaKGWK13})
  including
  \Term{confirmation bias} 
  (Munaf\`o \etal 2017\nocite{MunMNB17})
  focussing on the effects that
  appear most significant and plausible; and
\item
  \Term{publication bias}: journals and their readers are interested
  in significant effects and therefore, these get published and achieve
  attention, even if they were the result of chance and correspond to the
  testing error of the first kind.
\end{itemize}
They all reflect a way of selecting, among several potential results, those
that are statistically most significant. Therefore, those effects 
that are estimated, by chance, stronger than their true values are, will
have a larger likelihood to be documented in publications. 
Thus, effects in publications of ``original studies'' tend to be larger
(in absolute value) than the potential true effects they claim to estimate,
and consequently, their nominal statistical significance is
(much) more pronounced than it should be.
Clearly, this bias is enhanced for studies with low statistical power
(\cite{vZwEC21}).

The high need for new topics for publications, including Ph.D.\ projects,
has lead to an inflation of studies that do not examine clearly relevant
and plausible scientific questions.
The tactic is to collect some data and search for patterns in many ways in
order to find any that appear statistically significant.
While such exploratory analyses may play a positive role in the advancement
of science and lead to surprising discoveries, the effects found in this
manner suffer from selective reporting bias and therefore need independent
validation.
These reasons suggest that one should judge the chances of repliction to
depend heavily on the characteristics of an original publication,
and the replication crisis can be interpreted as the consequence of
unwarranted optimism (\cite{GibE21}; \cite{AmrVTG19}).

Comments on the crisis of replicability often complain about
mediocre quality, lack of education leading to inadequate experimental and
statistical procedures, and low statistical power, see
\citeasnoun{CleM15} for a review.
These aspects may lead to even more blind exploration entailing
higher selective reporting biases, but are otherwise not related to
replicability.

In the famous paper on ``Estimating the reproducibility of psychological
science'' (\cite{OpeSC15})
abbreviated as OSC15 in the sequel,
223 scientists were involved in replicating 100 statements about an effect
of some sort that had been published in high ranking psychological journals.
Among the 97 effects that had been statistically significant in the
original study, 35 reached significance in the same direction in the
replication.
This result was received as a shock, documenting the crisis in empirical
science. \citeasnoun{PatPPL16b} (and papers cited there) put the results into
perspective. 
In the meantime, there have been several similar attempts to obtain rates 
of successful replication, see
Klein \etal\nocite{KleRRV14} (2014, 2018); \nocite{KleRVH.18}
Ebersole \etal\nocite{EbeCMBB.20} (2020);
Camerer \etal\nocite{CamCDH18} (2018);
Cova \etal\nocite{CovFSAA18} (2018); and
Errington \etal\nocite{ErrTMS.21} (2021).

We do not discuss the important aspects of Good Practice in empirical
research in general here, see \citeasnoun{ShrPR18} and their summary
(their Table 1).
For broad discussions about reproducibility, mostly in the sense of
successful replication, see
\citeasnoun{ShrPR18}; \citeasnoun{BegCI15}; \citeasnoun{LeeJJ17}; \citeasnoun{MaxSLH15},
\citeasnoun{PatPPL16}; \citeasnoun{StaTCD18}; \citeasnoun{PleH18},
as well as the
``consensus report'' of the U.S.\
\citeasnoun{NatAcad19} with its broad view
on Best Practices in the scientific process.

\subsection{The generic case}

A generic problem consists of assessing the difference
between two groups of values of a continuous target variable,
the groups referring to different situations or treatments.
The difference is interpreted as the \Term{effect} of the
situation or treatment.
This problem will be used as the basic example
when introducing concepts and criteria below.

There are, of course, plenty of other ways in which data can lead to or
confirm knowledge. The majority of well-posed scientific problems leads to
measuring or observing a target variable (or sometimes more than one)
in different situations determined by ``explanatory'' or ``input''
variables. The interest is again in \Term{effects} of the latter variables
on the target.
Concepts and criteria should easily generalize to such situations.

While estimation of one or several effects is the central paradigm for
which reproducibility is widely discussed, other types of
statistical problems, like prediction, model development, or search for
potential causal relationships in big data, are neglected, 
see \citeasnoun{StaW16}.

\subsection{Focus of this paper}
A first topic of this paper is based on the observation that the
discussions have suffered from a lack of common language.
Even the term ``reproducibility'' has been used to name specific aspects
of validation.
And what does it mean to say ``This claim has not been reproduced''?
Was the analysis of the data misguided in the first place?
Has replication of the study not been possible or not been tried?
Have the results of a replication shown ambiguous
or even contradictory results?
Here, we try to establish clear terminology,
building on the different earlier proposals from different fields of research
and models of the scientific process. 

A second point is the assessment of success of a replication study.
Clearly, when such a study is undertaken to validate a claim
found in an empirical study,
one must expect the new data to lead to non-identical results due to
random variability.
When is such a replication successful?
In the literature, a popular but criticized criterion consists of finding
a statistically significant effect again.
Otherwise, the answer is often left to vague formulations such as 
``consistent effect sizes,''
``consistent measures of statistical significance,''
or ``the results [should be] within the range of values predicted by
estimates from the original study''
(see, e.g, \cite{NosBE17}).
\citeasnoun{AndSM16} distinguish between different
goals of replication and define the respective criteria for success. 
We will discuss and introduce precise notions and a classification of
results that goes beyond a simple binary answer in
Section~\ref{sec:assessment}.
The lack of criteria has lead to disputes about the interpretation
of replication studies, diagnosed as a ``war'' by 
\citeasnoun{IoaJ17}.

\label{sec:thispaper}
A fundamental issue is the perversion of scientific reasoning mentioned
above, consisting of a search for patterns in data and subsequently
treating the most salient ones by the statistical inference tools that are
adequate for testng pre-conceived hypotheses.
The road that leads to scientific knowledge starts from a clear scientific
question or hypothesis and then chooses the experiment or observation study
and the statistical tools to find the answer.
Now, a suitable question asks if there is a certain effect of interest.
We argue that since there is almost always at least a tiny effect,
the question needs to be enhanced by specifying a threshold of relevance.
This leads to shifting the focus away from testing to estimation
(Section \ref{sec:nhst}).
This issue concerns empirical studies in general, but also affects the
interpretation of replication results specifically.

Another fundamental issue in empirical science emerges from the experience
that there is generally a variability of results that goes beyond the
exptected statistical variation stemming from the randomness of the
single observations.
No replication study exactly mimiks the original, and assuming that
the new observations are independent realizations from precisely the same
distribution as the original ones is an over-simplification.
A reasonable model postulates a ``between study variance component'' in
addition to the variance of observations within the same study. It is
called the random effects model in meta-analysis (Section \ref{sec:between}). 
As a consequence, more than one replication is needed to assess the precision
of an effect estimate adequately.

These considerations show the complexity of the basic and seemingly
simple task of assessing an effect.
The final sections \ref{sec:repl} and \ref{sec:concl}
add thoughts about ways to overcome the difficulties.


\section{Terminology}
\label{sec:terms}
As mentioned above, terms used in discussions about ``reproducibility'' are
sometimes unclear or ambiguous.
Here, we collect them in best agreement with the literature as far as
there is a widespread common understanding and mention alternative
meanings. Words in \textit{slanted font} are meant to be used
as terms with the given meaning.

\Tit{Confirmability.}
The term \Term{reproducibility} formerly used to concern
the general requirement for scientific statements to be
confirmed upon repeating the research that has lead to them.
Since in the last decades, the literature converged to using
the term in a narrow sense, see below, we propose the term
Term{confirmability} instead as the \textsl{name of the theme}:
it encompasses all aspects of examination of the reliability
and relevance of a \Term{scientific claim} or \Term{statement}.

As mentioned in the Introduction, we focus on the situation where
a scientific claim resulting from an \Term{original study} is examined by
conducting a \Term{validation study}.
An empirical study typically consists of the following steps,
which may or may not be copied as far as possible in the validation study
(see also the Supplement of \cite{PatPPL16}).
\begin{enumerate}
\item
  Specification of the scientific hypothesis or \Term{claim} with a
  supposed domain of validity
  (like population, conditions, ranges of specified variables).
  Often, a replication only picks up a part of the original study.
\item
  Design of an experiment or specification of observation units,
  like subjects, animals, plots.
\item
  Tools: Auxiliary material, measurement devices, experimenter or observer.
  These may show batch or calibration effects, temporal variations,
  and environmental influences.
\item
  Generation of the data.
\item
  Data cleaning.
\item
  Statistical model and procedure of analysis.
\item
  Selection and presentation of results.
\item
  Interpretation.
\end{enumerate}

\Tit{Transparency, Re-assessment.}
All these steps should be documented well enough to allow for copying them
as far as possible. This includes public availability of the data and the
code to repeat the formal results, typically the output of statistical
methods. We call this aspect the \Term{transparency} of the steps.
(
Nosek \etal\nocite{NosBHMA.21} (2021) 
label it \Term{process reproducibility}, and
Seibold \etal \nocite{SeiHCDD.21} (2021),
\Term{computational reproducibility.)}
The steps should be verified in peer \Term{re-assessments} of publications,
and the result should be positive---a \Term{confirming re-assessment}.
The actual \Term{re-computation} might also be called a
\Term{verification} (\cite{CleM15})
of the analysis.
 \citeasnoun{DreAJ24} distinguish \Term{computational reproducibility}
(\Term{verification test} in \citeasnoun{CleM15},
\Term{recreate reproducibility}, calculation of the same statistical method
with a different program, and \Term{robustness reproducibility}, if the
conclusions are unchanged by ``alternative reasonable analytical decisions''

The term ``reproducibility'' is usually used for these aspects,
leading to ``reproducible research''
(\cite{PenR11}; \cite{PenR09}).
Alternatively, we may call it \Term{transparent research}.

\begin{remark}
In fact, this restricted meaning has lead to detrimental confusion,
when the ``reproducibility crisis'' was interpreted as a problem
of mistakes or ambiguities in applying statistical software.
As mentioned above, his is the reason for proposing a new term for the
encompassing theme, \Term{confirmability}.
\end{remark}

\Tit{Re-analysis.}
Some steps from ``data'' to ``interpretation'' may be subject to criticism,
and alternatives may be available. Then, a \Term{re-analysis} may be
appropriate as a validation of the claim of the original study.
A \Term{new analysis} may use the data for reaching new claims in the
spirit of an exploratory study.

\rem{unjustified or wrong data analysis and/or conclusions
  should not be reproduced
}

\Tit{Replication.}
When the experiment or the observation campaign is done again to obtain
new data---that is, steps ``design'' to ``data generation'' are again
executed in the same way---the study is called a \Term{repetition}
if it is done by the same team with the same set of tools.
Such a repetition should usually be published as part of an original study.
If a different team sets up the same experiment or observes according to
the same plan, it leads to a (``independent'') \Term{replication},
more precisely, a \Term{direct} or \Term{close replication}.

The term \Term{replicability} is generally used in an unfortunate because
misleading way. The literal meaning clearly is the
\Term{feasibility of replication}, without any specification of the
result.
A statement like ``This paper (or claim) is not replicable''
might (and should) mean that there is not enough documentation or that the
nature of the phenomenon does not allow for a repetition
(like the Big Bang), but it usually means that there has been a replication
study that failed to find the same result.
One should therefore state if a replication of a claim is feasible and
whether the result is a \Term{confirmation}, a \Term{failed confirmation},
or even a \Term{contradiction}.
We come back to this assessment of the result in Section \ref{sec:assessment}
and in the conclusions.

\Tit{Robustness, generalization, extension.}
It may be informative to examine whether scientific conclusions remain unchanged
when experimental methods or schemes of observation are varied, or
alternative data cleaning and statistical methods are used.
This desired stability has been called \Term{robustness} by
\citeasnoun{GooSFI16}. Since this word has many meanings
throughout science, it is recommendable to specify the steps (2, 3, 5, 6)
towards which the validatoin is tuned.

\Skip{
  \citeasnoun{CleM15} write:
  ``The critical distinction between replication and robustness is whether or
  not the follow-up test should give, in expectation, the same quantitative
  result as the original test within rounding error.''
}

More broadly, the degree to which a scientific claim remains valid if
conditions and populations different from those in the original study
is of profound interest,
Corresponding studies modify steps ``design'' to ``data generation''
and are called \Term{generalization} studies.

Similarly, a \Term{conceptual replication} is a study to assess
the validity of a claim under different circumstances, for other
populations, or using alternative methods of measurement, thus varying
the first three steps.
Such conceptual replication has been practiced in psychology in order
to derive psychological ``constructs''
(like measures of dimensions of intelligence) and to confirm
relations between these using different questionnaires or tests.

An \Term{extension} study would also extend the scientific claim itself.

\Tit{Further literature about terminology.}
Patil \etal (2016) \nocite{PatPPL16} propose terms that are
mostly similar to ours.
They distinguish between a ``replicable study'' which means that
the estimates of parameters are compatible, and a ``replicable claim,'' for
which the scientific conclusion is confirmed by the replication.


\citeasnoun{GooSFI16} introduce a distinction between
\begin{itemize}
\item 
  ``methods reproducibility,'' which
  corresponds to our ``transparency'' and to reproducibility as used
  in computational sciences,
\item
  ``results reproducibility,'' which corresponds to confirming
  replication, and
\item
  ``inferential reproducibility,'' defined as
  ``making knowledge claims of similar strength from a study
  replication or reanalysis.''
\end{itemize}

For further elaborations on terminology, see \citeasnoun{PatPPL16},
\citeasnoun{PleH18}; \citeasnoun{GooSFI16}; \citeasnoun{NatAcad19}, and
\citeasnoun{CleM15}.

\section{Estimation, not ``zero hypothesis testing''}
\label{sec:nhst}
It is common practice in most fields of empirical research to report
effects if and only if they are statistically significantly different
from zero.
This habit of ``Null Hypothesis Statistical Testing (NHST)'' has been
critized since it has emerged in the literature, but has lead to more
intense controversy in the last decades.
Since the p-value is usually given as a summary of the test's outcome,
the discussion is also named the ``p-value debate.''
The American Statistical Association has issued a broadly supported
statement against the unqualified use of this value (\cite{WasR16},
which however has not had strong impact on statistical practice yet
(\cite{GooS19}).

In a preceding paper (\cite{StaW21}),
this author has treated
the issue in depth and suggested a simple way to deal with it.
Here is a summary.

To fix ideas, consider a paired sample study, in which two
treatments are applied to each of $n$ observation units.
The effect is measured by the average of the $n$ differences $D_i$
of a target variable 
for the two treatments.
The parameter of interest, $\theta$, is the expected value of the $D_i$'s.
The classical test for the null hypothesis of zero difference is the
paired samples t-test, that is, the one-sample t-test on the differences
$D_i$.
The confidence interval is the t-interval corresponding to this test.

\begin{remark}
Note that in practice, it is preferable to use the Wilcoxon signed rank
test and the corresponding confidence interval.
The t test and interval are nevertheless still more often used, and they
generalize to other situations easily.
\end{remark}

In a general case, there is a model for $n$ observations $Y_i$,
containing a parameter $\theta$ of interest.
In most cases, a suitable estimator $\wh\theta$,
like the maximum likelihood estimator,
follows aprroximately a normal distribution,
$\wh\theta\sim\N\fn{\theta, V/n}$, where $V$ is the asymptotic variance.

The essential argument against NHST runs as follows
(see also \cite{GelAHY09}). 

\Tit{The Zero Hypothesis Testing Paradox.}
Testing an effect against zero does not answer a scientifically
meaningful question. 
When a study is undertaken to find some difference between groups or some 
influence between variables, the \emph{true} effect $\theta$
will never be precisely zero. 
Therefore, the strawman {null} hypothesis of zero
{true} effect {(the ``zero hypothesis'')} could
in almost all reasonable applications be rejected if one had the
patience and resources to obtain enough observations.
{Consequently}, the question that is answered mutates to:
``Did we produce {sufficiently many} observations to 'prove'
the (alternative) hypothesis that was true on an apriori basis?''
This does not seem to be a fascinating task.

This paradox has been stated prominently as
  a problem in the philosophy of science over fifty years ago
  in a highly cited long paper by \citeasnoun{MeeP67}.

\begin{remark}
  Researchers have taken the paradox into account by refraining from
  ``too large'' samples, thereby avoiding that tiny effects become
  significant.
  This pragmatic behavior nevertheless appears difficult to justify
  rationally.
\end{remark}

\Tit{Parametric models.}
The scientifically justified question is therefore:
``How large is the effect?''
The question makes sense only if the parameter is part of a model that
describes the phenomenon under study.
It can be asked independently of a design of an experiment or obbervation
scheme that is used to provide an answer.

\Tit{Estimation!}
The straightforward answer to the question is given by an estimate
with a confidence interval, based on data related to the model.
Many authors and teachers have propageted the routine use of confidence
intervals for statistical inference,
but the magic of expressing a result in just a single, scaleless
number---the p-value---has won in practice.
Here, we need to note a different limitation:
If effects are reported just when their confidence intervals do
not include zero, the selection bias still operates, and this problem of
the zero hypothesis testing ``culture'' is not avoided.

\Tit{Relevance threshold.}
In view of the Zero Hypothesis Testing Paradox, the scientifically
sensible question is:
``Is the effect relevant?''
This question asks for a \Term{threshold of relevance,} to be set by
informed judgement.
The threshold has been labelled
``Smallest Effect Size Of Interest (SESOI)'' (\cite{HeiR19}), 
``Minimum Practically Significant Distance (MPSD)'' (\cite{GooWSK19})
or the limit of the ``Region of Practical Equivalence (ROPE)''
(\cite{KruJ18}).

\Tit{Effect scale.}
The choice of a relevance threshold is often eased by expressing the effect
of interest on a natural scale, resulting from a transformation of the
original effect parameter.
Two widespread situations in pratice are the following.

\textsl{Proportions.} \
Distinctions are often expressed naturally as percentages.
Then, inference should be based on parameters and data transformed to
logarithmic scale. This translates percentage differences and
multiplicative effects into linear differences and effects,
which are simpler to interpret and treat mathematically.
Specifically, a chosen relevance threshold on the log scale corresponds to
a threshold on a multiplicative effect on the original scale.

\textsl{Probabilities.} \
For comparison of  probabilities, the log-odds or logistic scale
turns effects on odds into linear differences.
Whereas a difference of propabilities of $0.1$ has a very different
importance depending on the two values---changing $0.5$ to $0.6$
is much less severe than changing $0.88$ to $0.98$---equal differences in 
log-odds can be interpreted as being equally relevant.
A relevance threshold for log-odds relates to a threshold on
multiplicative changes of odds and identifies changes in probabilities
that are (arguably) intuitively comparable regardless of their numerical
values---a change from $0.5$ to $0.6$ appears equivalent to a move from
$0.88$ to $0.92$.

An effect scale is suitable if equal differences on it correspond naturally
to equally important effects on the original scale, and a constant
relevance threshold therefore applies to the whole range of possible
values. 
The transformed parameter will be called the effect $\eff$.

\textsl{Standardized difference.}
For many quantitative target variables, equal differences
on their original scale correspond to equally important effects, and
no transformation is called for. However, it makes sense to compare an
effect to the variable's random variability between observations.
In the generic case of a paired sample, this standardization amounts
to dividing the expected mean $\theta$ by
the standard deviation $\sigma$ of the distribution of the $D_i$'s
to get the standardized difference $\delta=\theta/\sigma$.
The standardized effect shall be $\eff=\delta/2$ for the sake of
consistency with the standardized coefficient in regression, see
\citeasnoun{StaW21}.

\Tit{Choice of a relevance threshold.}
The choice of a threshold may appear like an undesirable burden
for the researcher and a source of arbitrariness.
The Zero Hypothesis Testing Paradox suggests that avoiding it is
the source of irrelevant or misleading research---certainly a worse
option. 

In order to alleviate the burden, \citeasnoun{StaW21}
gives advice for a ``default'' choice for the most commonly used
statistical models.
If the logarithmic scale is appropriate, a threshold of $0.1$,
corresponding to a descrepancy of approximately $10\%$
on the original scale, is proposed as a plausible choice.
An analogous choice applies to the logistic scale, which is suitable for
proportions.
For a standardized effect, the recommendation is again~$0.1$.
The choice of these defaults has no deep scientific basis, but is arbitrary
to a similar degree as is the value of $5\%$ for the level of significance
tests. Any alternative with better justification in a specific context is
preferable. 

\begin{remark}
  Effect scales thus also make effects on the various corresponding types
  of target variables comparable.
  In many replicability studies, effects have been aligned by
  transforming them into correlations, following OSC15.
  However, effects on this scale are comparable only since they are centered
  such that the null effect turns into zero correlation, and transformed
  effects are bound by $-1$ and $1$.
  According to the arguments above, the correlation scale is not a suitable
  effect scale.
\end{remark}

Usually, the effect is supposed to be in one of the two possible
directions. In order to simplify the wording, we assume it to be positive
in the rest of the section.

\Tit{Relevance measure.}
Relevance can be expressed as the effect $\eff$, divided by the
relevance threshold $\zeta$, $\Rl{}=\eff/\zeta$.
Then, a value of $\Rl{}$ larger than 1 indicates a relevant result.
It is a parameter of the model, and the point and interval estimates
for the original model parameter that determines the effect leads to
an estimate $\Rl e$ and a confidence interval 
for the relevance. The lower and upper ends of the interval are called
the ``secured'' relevance $\Rl s$ and
the ``potential'' relevance $\Rl p$, respectively.
If the secured relevance is larger than $1$, the effect is statistically
proven to be relevant in a clearly defined sense.
Thus, \Rl s can be used as a new single number to summarize the most
important aspect of inference---what the p-value was meant to accomplish.

\Tit{Classification of results.}
Based on relevance, a differential answer to the research question
can be given according to the following distinction.
\label{sec:classif}
\begin{itemize} 
\item[Rlv]
  The effect is clearly relevant if
  the whole confidence interval is larger than the threshold, $\Rl s\ge 1$.
\item[Ngl]
  The effect is clearly irrelevant or negligible if
  the whole confidence interval lies on the low side of the
  threshold, $\Rl p<1$---whether or not zero is covered, that is,
  the null hypothesis is rejected.
\item[Ctr]
  The assumed direction of the effect proves wrong if the whole confidence
  interval lies on the negative side, $\Rl p<0$,
  a clear contradiction.
\item[Amb]
  The result is ambiguous if relevance 1 is contained in the
  confidence interval, $\Rl s<1<\Rl p$.
\item[Amb.Sig]
  It may be worthwile to label the sub-case of \ Amb \ in which the result
  is at least significantly larger than 0, $\Rl s>0$, as \ Amb.Sig.
\item[Ngl.Sig]
  Similarly, the sub-case of \ Ngl \ with a significant effect is \ Ngl.Sig.
  It will be very rare unless the sample size or the relevance threshold
  is large.
\end{itemize}

\unitlength 0.9mm
\def\Label#1{\llap{\textbf{#1}:}\ }

\if y\Figures
  \includegraphics[width=\textwidth]{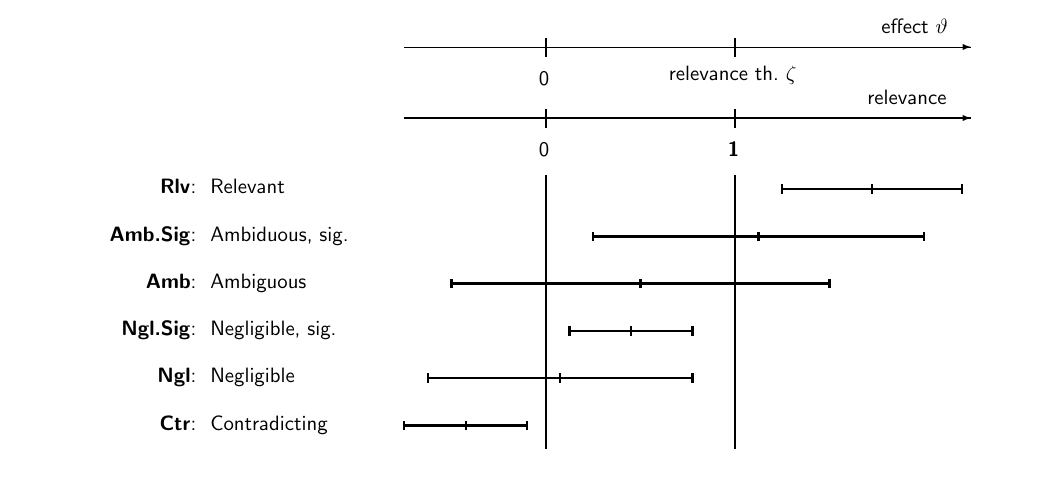}\fi

\begin{figure}[h]
  \caption{Classification of cases based on a confidence interval and a
    relevance threshold}
  \label{Fig1}
\end{figure}

The problem of Zero Hypothesis Testing has often been mentioned as
an important cause of the reproducibility crisis.
In fact, the relation between the two themes is only indirect:
The testing routine leads to efficient screening among
multiple possible effects for the most appearently significant ones
and thereby entails the selective reporting bias, which is, as stated
above, the direct reason for repliclation failure.
  
\begin{remark}
  Bayesian inference has been advocated as the fruitful alternative to
  NHST. An important argument is the possibility to assign and develop
  a probability for the null hypothesis to be true.
  If the null hypothesis consists of a zero effect, our argument says
  that its probability is zero, which contradicts this idea.
  For a ``fat'' null hypothesis, Bayesian analysis makes sense,
  but also needs a relevance threshold, and additionally, to the
  specification of a prior distribution on the effect,
  see \citeasnoun{KruJ18}.
  Note that our terminology may be adapted to any versions of Basian point
  estimates and credible intervals.
\end{remark}

\section{Assessment of Success}
\label{sec:assessment}
Let us come back to the theme of replicating an ``original'' study that
examined an effect. 
How should we assess the success of the replication?

There are two main aspects of success or failure:
\begin{itemize}
\item
  Does the replication lead to the same conclusion as the original?
  See \ref{sec:sigag}.
\item
  Are the two studies consistent in the sense that their quantitative
  results are similar?
  See \ref{sec:consistency} to \ref{sec:between}.
\end{itemize}

\subsection{``Significant again''-- or ``Relevant Again''}
\label{sec:sigag}
The hypothesis testing mode of reasoning suggests that the replication
of a significant effect is successful if it turns out significant again
and the estimate has the same sign in both studies.
Let us call this criterion ``significant again,'' ``sigag.''

In the rare case when an insignificant original effect is examined by
replication, success would be to reach insignificance again,
but a study with this goal would be, one could say, of ``null merit,''
c.f.\ the Zero Hypothesis Testing Paradox above, and
\citeasnoun{PawSHMH24}.
A meaningful question in this case may ask if the effect can be shown to be
negligible, see Section \ref{sec:nhst}.

In the light of the preceding section, ``sigag'' is not a sensible
criterion. 
Note that the probability of achieving it depends on the p-value of the
original result. Assume for a moment that the original result was just
significant (p=0.05) and the original estimate happened
to be the true value of the effect.
Then (neglecting that the scale parameter needs to be re-estimated),
obtaining a larger estimated effect in the replication with the same number
of observations amounts to ``sigag'' and has a probability of about 0.5
(\cite{GooS92} and
Piper \etal\nocite{PipSGRR19} 2019).

In OSC15, the proportion of studies achieving ``sigag'' was 36\%.

One can ask for the same conclusion again also using relevance:
A relevant result in the original study (case Rlv) is successfully
replicated if the case Rlv shows up again, leading to
``relevant again.''
Clearly, in the respective borderline case, this has again
a probability of 0.5.
We will come back to this requirement and call it a ``confirmation''
when we distinguish more cases than
just ``success'' from ``failure'' in Section~\ref{sec:ass.class}.

\subsection{Consistency}
\label{sec:consistency}
In empirical studies, the data is subject to random variation.
This applies to the original as well as to the replication study.
A reasonable question to ask is whether the data in the two studies
could be described as coming from the same statistical population.
In the generic case, this can be checked by testing if the average
$D_i$ in the original study and in the replication show a
statistically significant difference.
The question is answered by a two-sample t-test.
(Note that, again, a nonparametric rank sum test would be more appropriate.)
If the test shows no significance, one can say that
\Term{the two samples are consistent.}

A closely related approach calculates a prediction interval
for the estimate of the replication from the estimate of the original
study, its precision, and the sample size of the replication,
and checks if the actual estimate of the replication falls into this
interval.
An equivalent criteron proceeds with inverted roles: the original value
should be contained its the prediction interval obtained from the
replication study. Both versions are equivalent to the t test mentioned
above (as long as the variance is estimated from both studies).
See also \citeasnoun{PatPPL16}; \citeasnoun{PawSH20}; and
\citeasnoun{SchaJH21}.
Applying the first version to OSC15,
75\% of the replications produced consistent data (\cite{PatPPL16}).
(
\citeasnoun{SchaJH21} present calculations of
the probability of successful replication if the true effects coincide.
Unfortunately, they were based on unconditional probabailities for both
studies rather than conditional ones, given a significant result in
the original study.)

\begin{remark}
Note that this approach does not compare the data in all aspects,
but concentrates on the effects.
In our generic case, the response values in the replication
could be quite divergent from those in the original study.
The test only checks the differences $D_i$ between the responses for
the two treatments.
If the response shows different values, the new study indeed examines
a generalization of the original results.
\end{remark}

It is tempting to say that success of the replication is achieved if
the confidence interval of the replication overlaps with the interval
of the original study.
A second thought shows that this criterion accepts compatibility more
often than is likely assumed by most readers:
Under the null hypothesis of equal effects and sample sizes in the
two studies, the probability of such an overlap is about 99.7\%
(based on the normal instead of a t distribution,
$\Phi(2\cdot 1.96/\sqrt 2)$)
instead of 95\%.

\begin{remark}
The OCS15 study used confidence intervals in an inappropriate way:
The replication was labelled as successful when the confidence interval
of the replication covered the \textit{estimated} effect of the original.
This criterion does not consider the randomness of the original result.
It is easy to see that if the power of the replication study was increased
sufficiently, the criterion woud almost certainly fail, regardless of
the quality of the original study. 
A symmetrized version checks ``whether the estimates are within each
other’s confidence intervals'' (\cite{ErrTIG14}) and suffers from the
same flaw.
In spite of these undesirable properties, these criteria are still in
use, see \citeasnoun{ErrTMS.21}. 
\end{remark}

\subsection{Relevant Effect difference}
Since we do not want to fall back on testing a hypothesis,
we now re-formulate the problem.
In the generic case, the quantity to be estimated is 
\textit{half the difference between the true treatment effects}---or 
more generally of a parameter in a given model---in the two studies,
$\ED=(\theta\sups r-\theta\sups o)/2$.
(The reason for choosing \emph{half} the difference is mentioned above.)

In order to ease interpretation and avoid cumbersome details, assume
that the effect in the original study is positive, $\theta\sups o>0$.
The typical case of an attenuated effect then leads to a negative $\ED$. 

The (approximate) confidence interval for \ED\ is determined by the
standard error $\seED\;$ obtained from the standard errors $\se\sups s$
of the effect estimates in the two studies,
\[
  \EDh \pm q \seED \;,\quad  
  \seED^2 = \big((\se\sups o)^2+(\se\sups r)^2\big)/4
  \;,
\]
where $q$ is the appropriate quantile of a t distribution.

As discussed in 
the previous section, the result should be interpreted
with reference to a threshold of relevance.
Since the plausible and relevant direction of the difference is to the
negative side (a smaller effect in the replication than in the original),
the threshold is applied with a minus sign.
Then, the case ``relevant (Rlv)'' occurs if the confidence interval
for $\ED$ lies on the low side of this threshold, and analogously for the
other cases of the classification in Section \ref{sec:classif}.

\Tit{Standardized Effect Difference, EDS.}
Apart from this, the considerations on selecting an effect scale
apply to the comparison of the replication with the original again.
The possibly transformed or standardized effect difference is called
$\EDS$.
The relevance threshold for the comparison may be chosen differently
from the threshold used for expressing the relevance of the effects
in the two studies.

Note that \EDS\ is a \textit{parameter of the model}. 
It is estimated by plugging in estimates of the parameters.
In our generic case, it is plausible to use the standardization
$\EDS=(\theta\sups r-\theta\sups o)\big/(2\sigma)
=\eff\sups r-\eff\sups o$,
where $\sigma$ is the standard deviation of the $D_i$s and is assumed
to be the same in the two studies.

\begin{remark}
  \label{rem:standCohen}
In the generic case, $2\,\EDSh$ equals an index that is
well-known in the social sciences, called Cohen's $d$ and is sometimes
called \Term{Standardized Mean Difference} in other sciences.
Note, however, that the index here refers to the difference between
studies, not as usual to the difference between groups within study.
\\
In fact, in many replication studies, Cohen's $d$ between groups
has been used as the effect size and calculated for both the original and
the replication study.
It is misleading to compare the ``$d$'s'' between the studies.
A difference,
$d\sups o-d\sups r =
\wb D\sups o/\wh\sigma\sups o - \wb D\sups r/\wh\sigma\sups r$,
could easily occur if the unstandardized effects $\wb D\sups s$
were equal in the two studies, but the variabilities $\wh\sigma\sups o$
and $\wh\sigma\sups r$ of the observations were different.
Such a difference between variabilities could be due to a true difference
in precision of the measurements or to chance.
Thus, the effect $\theta$ itself must not be standardized when
compared between studies, but the standaridization of their difference by a
common scaling parameter $\sigma$ is appropriate.
The same argument applies to some other transformations of effects, like
the transformation to a correlation coefficient applied in OSC15 and
other replication campaigns.
\\
We emphasize this point because it has not been taken into account even
in the discussion of the likelihood of successful replication by
\citeasnoun{SchaJH21}.
\end{remark}

In the generic case, $\EDS$ is therefore estimated by
\[
  \EDSh = (\wh\eff\sups r-\wh\eff\sups o)/\wh\sigma
  =\wh\eff_p\sups r-\wh\eff_p\sups o
  \;,
\]
where $\wh\sigma$ is the pooled estimate of $\sigma$,
and thus $\wh\eff_p\sups o=\wh\theta\sups o/\wh\sigma$
(subscript $p$ for ``pooled'')
differs from the standardized effect in the original study,
$\wh\eff\sups o = \wh\theta\sups o/\wh\sigma\sups o$, and analogously for
the estimates in the replication.
$\EDSh$ is then proportional to the t test statistic 
$T$ for comparison of two independent samples, $T = 2\,\EDSh/c_n$
with 
$c_n=\sqrt{1/n\sups o+1/n\sups r}$.
The confidence interval for $2\,\EDS$ is thus given by
$  2\,\EDSh \pm q\cdot c_n $,
where $q$ is 
the quantile of the t distribution with $n\sups o+n\sups r-2$ \
degrees of freedom. 

A threshold of $0.1$ for \EDS\ equals the threshold $0.2$
for ``small'' values for Cohen's standardized difference $d=2\,\EDS$
that is popular when interpreting $d$.

\Tit{Standardization in the general case.}
In a general setup, the ``effect'' $\theta$ is any parameter
of interest in a given model describing the observations.
An estimator $\wh\theta$ has a given distribution, derived from the model.
Usually, this distribution approximately equals a Gaussian with
a variance that is inversely proportional to the sample size,
$\mbox{var}(\wh\theta) = V/n$, $\wh\theta\sim\N(\vartheta,V/n)$.
Then, the estimator of $2\,\ED=\theta\sups r-\theta\sups o$ entails the
confidence interval 
\[
  \big(\wh\theta\sups r-\wh\theta\sups o\big) \pm q\;
  \sqrt{\wh V\sups o/n\sups o+\wh V\sups r/n\sups r}
  \;.
\]
Often, $V$ does not depend on the value of $\theta$,
or this can be achieved by a transformation of $\theta$.
Then, the standardized effect difference is
\[
  \EDS = \big(\theta\sups r-\theta\sups o\big)\Big/
  \big(2\sqrt {V}\big)
  \;.\]

Note that standardization is not needed nor recommended if the effect scale
is logarithmic or logistic.

\subsection{Desirable properties of discrepancy measures.}
The quantity \ \EDS\ 
has the first three of the following properties that we consider
essential for any index \ RD \ of
\Term{\textit{R}eplication \textit{D}iscrepancy}. 
\begin{itemize}
\item[(P1)]
  RD should be a function of parameters of the model for the original
  and the replication(s) and thus should not depend on the number of
  observations used in the studies. 
  It is then \textit{estimated} on the basis of the data.
  (See ``Parametric models'' in Section \ref{sec:nhst}.)
\item[(P2)]
  RD should measure the discrepancy between the quantities of interest
  in the original and the replication study (or studies) and be
  rather insensitive to differences in other aspects.
\item[(P3)]
  It is desirable that RD can be generalized to multivariate effects,
  as for instance to analysis of variance, where several contrasts
  are of interest.
\item[(P4)]
  RD should generalize to the situation of more than one replication
  study.
\end{itemize}
The desired property (P4) leads us to extending our model as follows.

\subsection{Heterogeneity of studies}
\label{sec:between}
Following the preceding arguments, one might expect that
the test for zero difference $\ED=0$ or $\EDS=0$ should fail only in 
about 5 percent of replications of original studies that are
free of selective reporting bias.
General experimental-statistical experience dampens this hope.
The hypothesis of exactly equal expected effects,
$\vartheta\sups r=\vartheta\sups o$, is not realistic in practice.
It is clear from experience of any types of measurements or observations
that their random variation within the same study will be smaller
than the variation of measurements from different studies.
In technical terms, there is a variance component reflecting
the differences between studies, the \Term{between studies variance}.
The concept is usually called ``heterogeneity'' and forms the basis of the
random effects model in meta-analysis.
In the generic case, the effects $\vartheta\sss$ in different
studies $s$ are modelled as realizations of a random variable.
The quantity of interest would be the expected value $\Theta$ of this
random variable, and its variance is the
``between study variance component'' $\sigma_\vartheta^2$.
Estimation of $\Theta$ is best achieved by an average
(possibly a weighted one) of the $\wh\vartheta\sss$ that are available,
and the width of a confidence interval would need to contain an
estimate of $\sigma_\vartheta$ or of the (relative)
\Term{Between Study Variability} $\mbox{BSV}=\sigma_\vartheta/\sigma$,

\citeasnoun{HedLS21} use this model as a basis for recommendations on
designing replication studies.
Heterogeneity has received increasing attention in recent years,
see, e.g, \citeasnoun{KenDJ19}, 
\citeasnoun{StaTCD18};
\citeasnoun{HedLS19}; and
\citeasnoun{MatMV19}. 
\nocite{HedLS19b}


\begin{remark}
  In meta-analysis, an index ($H$ in \cite{HigJT02}) compares a version
  of the between-study variability $\sigma_\vartheta^2$ with the average of
  precisions $1/\var\big({\wh\vartheta\sss}\big)=n\sss/V\sss$ of the effect estimates
  for the individual studies.
  In contrast, BSV uses the average of the $V\sss$ quantities, which describe
  the information contained in individual observations rather than the
  variability of the estimates.
  This makes BSV a parameter of the model for the observations that is
  independent of the numbers $n\sss$ of observations in the studies,
  thus fulfilling property (P1), whereas
  the random effects model of meta-analysis starts from the
  effect estimates in the studies and their precisions and therefore fails
  to characterize the basic phenomenon generating the data.
\end{remark}

\begin{remark}
  \citeasnoun{CleM15} states that 
``In expectation, these tests [the tests for 0 difference of effect sizes]
are supposed to yield estimates identical
to the original study. If they do not, then either the original or
the replication contains a fluke, a mistake, or fraud.''
In the light of the concept of a between study
variance component, such a conclusion is not warranted.
Several authors suggest potential reasons for heterogeneity and urge
researchers to investigate and eliminate them.
Experience from interlaboratory studies in chemistry and metrology in
general shows that often, no reasons for a variance component between
batches can be identified, but it remains relevant anyway.
\end{remark}

If only the original and one replication study are available,
an estimate of the between studies variance component is only possible
if the selective reporting bias of the original is assumed to be zero,
and it then relies on one degree of freedom
and would therefore be ill-determined and useless.
(In fact, $\BSVh^2=\big(\EDSh^2-(1/n_0+1/n_1)\big)/2$---or $0$,
if this is negative---can be interpreted as a point estimate
of the between study variability $\mbox{BSV}$.)

\Tit{Several replications!}
Remembering that a valid confidence interval
for the true effect $\Theta$ needs a reliable value for the
between study variance $\sigma_\vartheta^2$,
a reasonable number of replications is needed for its estimation,
as pointed out and justified by 
\citeasnoun{HedLS19}.
As Benjamini (2020, Section 6) \nocite{BenY20} writes:
``Replication results can and should accumulate, and for that purpose
many small studies are generally better than a single large one of
the same total size, even if each small study is underpowered,
because they reflect the relevant study to study variability.''
Due to the selective reporting bias, the original study should not be used
in the inference about the parameter (\cite{PawSH20}).

\Tit{Prior knowledge.}
Alternatively, if a number of replication studies for different
original claims in a field of application should lead to
similar values of \BSV,
such a value may be used to calculate a rough factor by which
the confidence interval for $\vartheta\sss$ of a single replication study
should be widened in order to be used as a confidence interval for the
``global'' true effect $\Theta$.

For a general Bayesian model that incorporates heterogeneity as well as prior
knowledge about the effect, including assumptions about selection bias,
see \citeasnoun{PawSCH23}.

\subsection{A classification of outcomes}
\label{sec:ass.class}
The goal of replication is to validate a scientific claim.
Here, we deal with the case of an ``effect'' that has been found
relevant or at least significant in the original study.
On the basis of the confidence interval ``$\IEff$'' for the
effect $\vartheta$ obtained in the replication and on the
confidence interval ``\IEDS'' for \ EDS, the result may be characterized,
using the scheme of Section \ref{sec:nhst}. Besides a threshold of relevance
for the effect $\vartheta$, a threshold for relevant values of the 
standardized effect difference EDS is needed.
EDS is relevant if it is lower than the negative relevance threshold.
Then, the result is a 
\begin{itemize}
\item[(Cnf)] 
  \Term{Confirmation,} if $\IEff$\  only contains relevant
  values (case Rlv), and
  the negative standardized effect difference EDS is small (cases Ngl or Amb);
  if $\IEff$\ is only significant (Amb.Sig) and
  the estimate $\wh\eff_1$ is larger than the relevance threshold,
  we call it a \Term{weak confirmation} (CnfW),
\item[(Att)]
  \Term{Attenuation,} if $\IEff$ \ lies on the same side of 0
  as in the original study (Rlv or Amb.Sig) and \IEDS\ is relevant (Rlv),
\item[(Enh)]
  \Term{Enhancement,} if the replication suggests a clearly
  stronger effect, that is, case (Rlv) for $\IEff$\
  and significantly positive EDS (Ctr);
  this will be rare,  
\item[(Amb)]
  \Term{Ambiguous,} if $\IEff$ \ covers the relevance threshold and
  it also covers zero (Amb) or the estimate $\wh\eff_1$ is below the
  reference threshold,
\item[(Anh)]
  \Term{Annihilation,} if $\IEff$\ covers only irrelevant values (Ngl), 
\item[(Ctr)]
  \Term{Contradiction,} if all values of $\IEff$\ have the
  opposite sign (Ctr),
\item[(Drp)]
  \Term{Dropout,} if the replication failed to mimik
  the experimental or observational setup.
\end{itemize}
The classification is summarized in Table \ref{tab:classification}
and displayed in Figure \ref{Fig2}.
The first three cases are identified by the ``significant again''
criterion as a \Term{successful replication.}
Nevertheless, conclusions might be rather different between them,
see Section~\ref{sec:strategy}.

\begin{table}[h]
  \centering
  \begin{tabular}{r|ccc}
    \multicolumn{1}{r||}{Effect estimate}
    &\multicolumn{3}{c}{Effect Difference (standardized), \IEDS}\\
    \multicolumn{1}{r||}{ $\IEff$\ in replication}
    & relevant, Rlv & Amb or Ngl & contradicting, Ctr \\\hline
    relevant, Rlv  & attenuation, Att & confirmation, Cnf & enhancement, Enh \\
    significant, Sig& attenuation, Att & weak conf., CnfW$^*$ & ---  \\
    ambiguous, Amb & ambiguous, Amb & ambiguous, Amb & --- \\
    negligible, Ngl& annihilation, Anh& annihilation, Anh$^{**}$ &  --- \\
    contradicting, Ctr& contradiction, Ctr&  --- & --- 
  \end{tabular}
  \caption{\label{tab:classification}Classification
    of results of a replication of a relevant effect, based on the
    classificaion of the confidence interval $\IEff$\ for the effect
    in the replication and the confidence interval \IEDS\ of the EDS.
    It is assumed that the original effect was relevant or at least significant.
    Then, the cases marked --- cannot occur.
    $^*$ This conclusion also requires $\Rl e\ge1$; otherwise, it counts as
    ambiguous. \
    $^{**}$ This cannot occur if the original effect was relevant.
  }
\end{table}

\if y\Figures
  \includegraphics[width=0.9\textwidth]{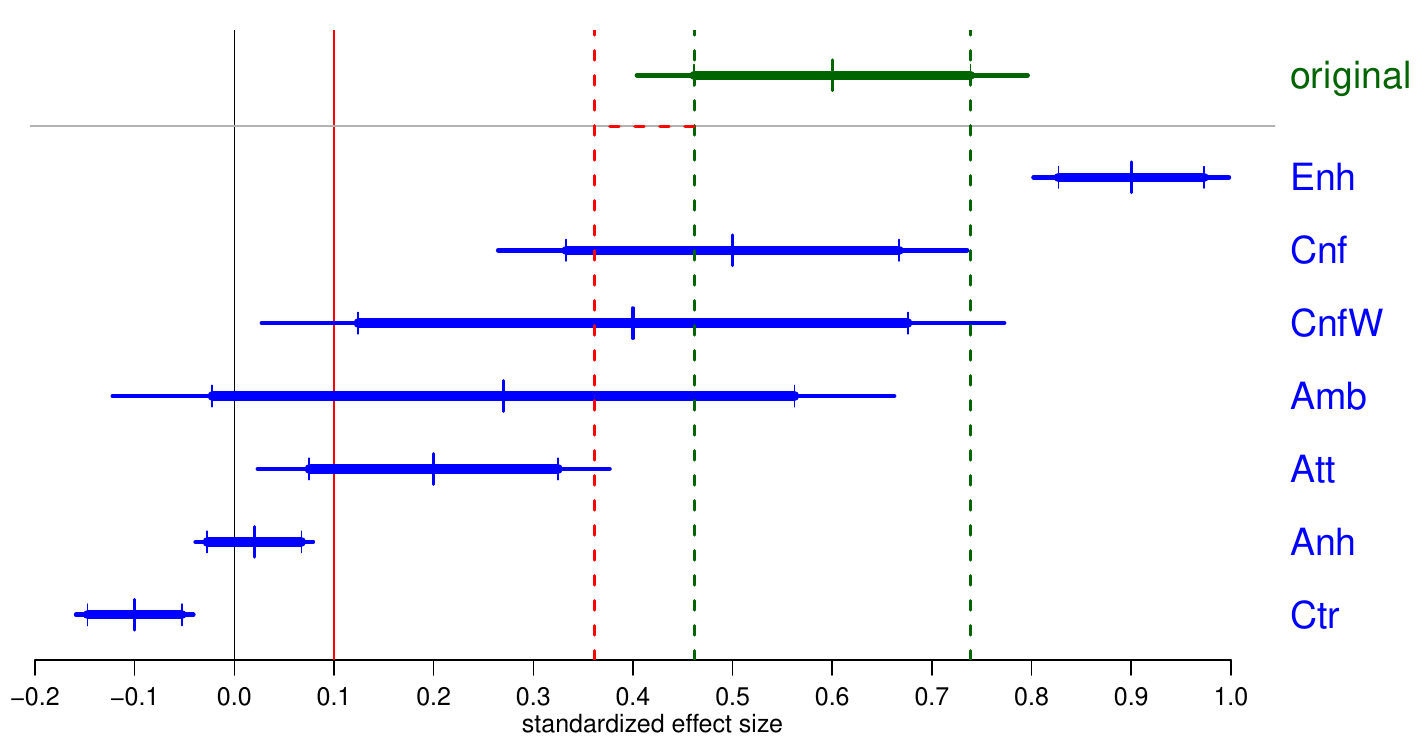}\fi
\begin{figure}[h]
  \caption{ Classification of replication results for a relevant original effect.
  Confidence intervals for the original effect (top, green) and for the
  effect obtained in the replication in the different possible cases
  (blue). 
  The additional ticks and thicker lines on the bars allow for an
  assessment of EDS. If the intervals bounded by them do not overlap,
  EDS is significantly different from 0.
  In order for EDS to be also relevant, the gap must be longer than
  $\rho_D=0.1$, say.
  Thus, for ``attenuation'', Att, the right hand end point of the
  bar's thicker part must be to the left of the red dashed vertical line.
}
\label{Fig2}
\end{figure}

\begin{remark}
  \citeasnoun{BonD20} also uses $\IEff$\ and \IEDS\ to classify replication
  results. His classes are defined by one of these intervals or the other.
  Therefore, they overlap. See supplementary material for more detail.
\end{remark}

As an illustration, Figure \ref{Fig3} shows
confidence intervals for standardized effects in ten ``simple cases''
studied in OSC15, together with their classification.
The data is displayed in the Supplement.
We suggest that showing the confidence intervals for the original study and
the replication(s), including the relevance threshold (cf.\ end of
Section \ref{sec:nhst}) should become a
standard display of the information in replication studies.
(Unfortunately, we had to compare standardized effects here, disregarding
Remark \ref{rem:standCohen}, since unstandardized effects are not given in
the data provided by OSC15.)

The figures display additional ticks on the interval bars that allow for
checking the significance of the effect difference $\ED$.
If the shortened intervals do not overlap, the difference is significant,
and the relative width of the gap or overlap visualizes how significant the
difference is indeed.
The position of the ticks is given, for the original study, by
$\wh\eff\sups o\pm q\nu\sups o\se\sups o$, where
$\nu\sups o = 2\,\seED\big/(\se\sups o+\se\sups r)$,
and analogously for the replication results.
Then, the gap is, if $\ED>0$,
\[
  \wh\eff\sups r - q\nu\sups r\se\sups r -
  (\wh\eff\sups o + q\nu\sups o\se\sups o)
  = 2\,\ED - (q\nu\sups o\se\sups o + q\nu\sups r\se\sups r)
  = 2\,(\ED - q\seED)
  \;,
\]
which is positive if and only if the difference is significant.
(This enhancement corresponds to the idea of ``notched box plots''
[\cite{McGRTL78}]
and makes it [asymptotically] exact for the comparison of two samples.)

\if y\Figures
  \includegraphics[width=0.9\textwidth]{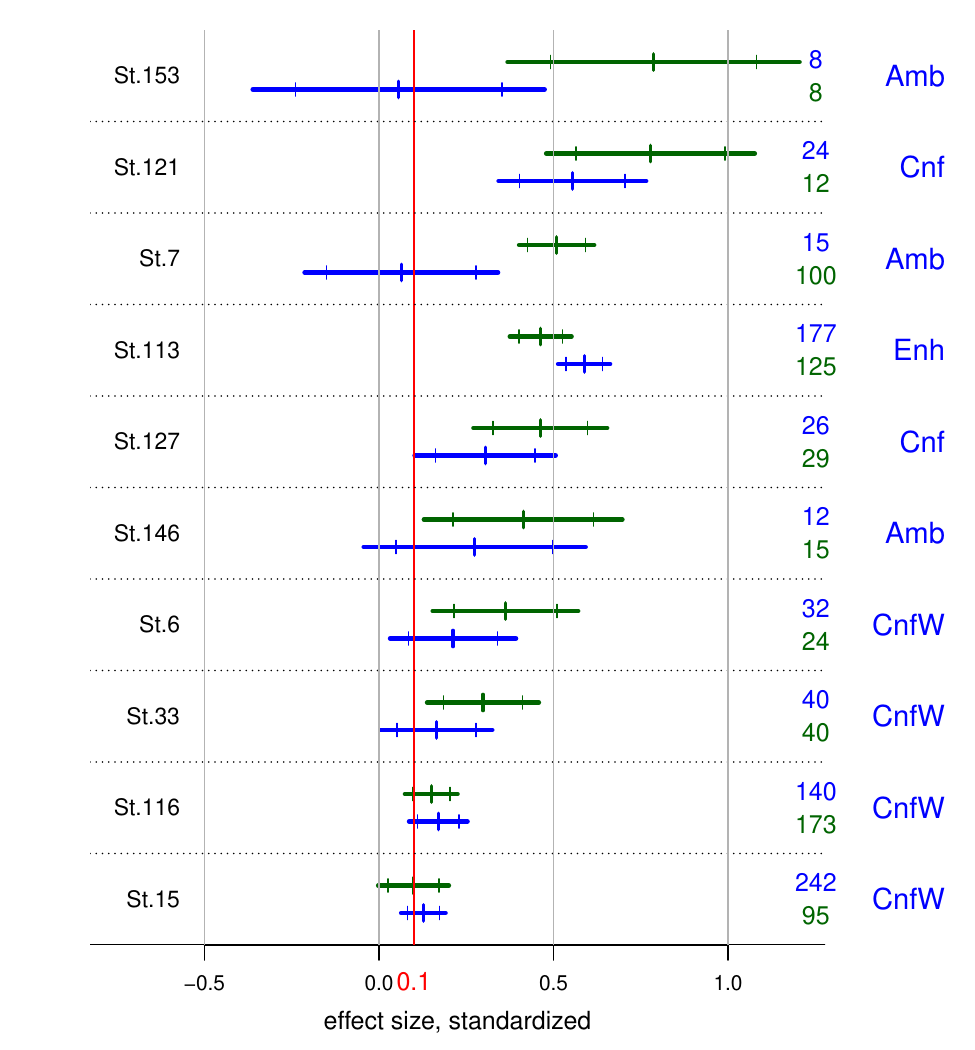}\fi
\begin{figure}[h]
  \caption{Confidence intervals for effects in ten items studied in OSC15,
  all based on paired or simple t-tests.
  Original and replication results are shown in green and blue,
  respectively. 
  The number of observations and the classification according to
  Table \ref{tab:classification} are shown in the right margin.
}
  \label{Fig3}
\end{figure}
\Tit{R package.}
The methods implementing the proposed ideas and exhibits are available in
the R package \Term{relevance} in the CRAN repository,
\texttt{cran.r-project.org}.
\section{Replication!}
\label{sec:repl}
Let us briefly come back to the different sources of selective reporting
bias.
It emerges in settings where data may show several or many effects or
patterns. 
By a formal or informal search, the most prominent ones are selected,
and statistical methods are applied that ignore this selection process.
Note that if a study is designed in the first place to find
``real'' effects among a clearly defined set of possible ones
in a structured way,
there often are statistical tools to avoid selective reporting bias,
notably the well-known Bonferroni rule, but
also more refined methods like those proposed in 
\citeasnoun{BenYH95}; \citeasnoun{MeiNMB09}; and
\citeasnoun{BerRBBZZ13}.
But in general, such methods to correct for selection are not available or
not used.

\Tit{Pre-registration!}
Publication bias is avoided by \Term{pre-registration}:
The scientific claim(s) to be examined and the plan to
perform a respective study are published
in an accessible and visible registry, before the data is collected.
Such \Term{pre-registration reports}---usually called
``registered reports''---should be peer-reviewed (\cite{ChaCT22}).
The results of the study must then be published regardless of whether
they confirm or contradict the claim.
Whereas such practice can be useful even for original work,
it must be standard for replication studies, since otherwise,
selection bias may easily play again and even a joint meta-analysis
of several replications could not be reliably interpreted.
\textit{A replication that is not pre-registered should not obtain the
  label ``replication,'' and its publication should be avoided,}
since this would entail the danger that it is included in a meta-analysis. 

\Tit{Several replications are needed!}
If a trustworthy confidence interval for the size of an effect is
  desired, the between studies variance component must be estimated.
  As argued above, several replications---at least 5, say,
  or 10, see \citeasnoun{MatMV20}---should be
  planned and pre-registered.

\begin{remark}
  Experience in the series of replications in cancer biology
  shows that in several studies, some of the preliminary steps needed 
  before acquiring the data for validating the original claims
  have failed and a section on ``Deviations from registered report''
  became necessary.
  When essential deviations occur, the replication should be counted
  as a ``dropout.''
\end{remark}

\begin{remark}
  A confidence interval for the effect, obtained in a pre-registered study,
  would have the correct probability of covering the true parameter value
  if there was no between studies variance component, see above.
  Note, however, that such a study may in fact be a step in a stepwise
  strategy (Section \ref{sec:strategy}):
  If the replication fails to be significant again,
  there is an incentive to try it again by another pre-registered
  replication.
  If this was a clear strategy, a formal treatment as a sequential
  procedure would be appropriate to derive correct coverage probabilities.
\end{remark}
\begin{remark}
  Extending this consideration further makes it clear that
  the selective reporting bias of scientific claims can be greatly reduced by
  replication studies and strategies, but not eliminated, because
  claims that are not confirmed will be forgotten, and those that
  are will be retained. This again constitutes a selection and leads to
  a ``secondary selective reporting bias.''
  The flaw is reduced if replications are restricted to the most relevant
  claims. A blind routine of replicating as much as possible would
  increase the secondary selective reporting bias.             
\end{remark}
\begin{remark}
  When planning a single replication, it is essential to consider
  the power of the new study---as a low powered single replicaton
  will enhance the problems just mentioned.
  Using the estimated effect from the original study as the
  true effect for power analysis would be inappropriate because of the
  selective reporting bias that is to be assumed for the original work.
  \citeasnoun{AndSKM17} and \citeasnoun{BonD20}
  consider the uncertainty of the original estimate
  in addition to the bias, discuss the implications and provide a method
  to calculate more appropriate sample sizes for a replication.
  \citeasnoun{HedLS21} take heterogeneity into account
  and \citeasnoun{PawSCH23} provide a general model that
  comprises both heterogeneity and bias.
  \citeasnoun{MicCH22} treat the idea of sequential replication
  additionally. 
  Implications of statistical power are discussed extensively by
  \citeasnoun{MorRL16} under the title
  ``Why most of psychology is statistically unfalsifiable.''
\end{remark}

\Tit{Performing replications can be made attractive.}
Even though the benefits of replication studies are widely recognized,
many authors seem rather pessimistic since they judge such studies to be
unattractive. They state that researchers will not get the necessary
recognition and funds if they invest their time and resources into
such activities. 

However, we see two ways that lead to the desired studies:
\begin{itemize}
\item
  Beginning Ph.D.\ students need to learn and practice the methods
  of scientific studies in their field.
  They often work in directions that start from an existing publication.
  If they are asked to perform a replication of such a study at the start,
  this guarantees a first publication, which is counted towards the number
  needed to complete their thesis (\cite{EveJE15}; \cite{KocAO18}).
\item
  More generally, research often aims at a generalization or extention
  in the sense of Section~\ref{sec:terms}. Such projects should
  contain a (pre-registered!) replication as a first part
  (Morey \etal\nocite{MorRCE16} 2016; \cite{BonD12}).
\end{itemize}
\citeasnoun{ChaC19} gives good arguments for a change of culture
rewarding replications.

\Tit{Let's establish the rules!}
The following rules should be suitable for establishing the replication
paradigm:
\begin{itemize}
\item
  \textbf{The Pottery Barn Rule} (\cite{SriS12}).
  Journals should adopt the policy of accepting pre-registrations for 
  replication studies of the ``original'' papers that they have
  published.
  They entertain a publicly available list of these pre-registered
  projects, preferably integrated with the lists of other journals.
  They guarantee that the results of the replication will be published:
  A short version must appear in their main mode of publication, possibly
  leaving the documentation to an online ``supplementary'' part.
  This should allow journals and their readers to keep their enthusiasm for
  novel findings any yet promote the establishment of reliable knowledge.
  
  Adequate power for deciding about the relevance of the effect
  need \emph{not} be required in each replication, as several independent
  ``underpowered'' replications are more easily obtained and more useful
  than a large replication study, since such studies eventually allow
  for estimation of the inter-study variability.
  Note, however, that a conclusive joint evaluation is only warranted for
  a pre-planned series, since otherwise, the results of the first
  studies might influence the likelihood that later studies will be
  undertaken. 
\item
  Supervisors and funding agencies of beginning Ph.D.\ students ask that they
  start with a replication study, preferably of an original study from
  another research group. This would even enhance cooperation among groups.
  (Clearly, this principle cannot be followed in fields where experiments
  take long---more than a year, say.)
\item
  In addition, the principles of open science, i.e., 
  complete transparency, data availability and re-computability of analysis
  help to improve reproducibility.
  Here, the platforms of the Open Science Framework (osf.org)
  \prepcite{NosBAB15} 
  (\cite{FosED17}; \cite{NosBAB15})
  and Zenodo (zenodo.org) are a very useful resources, and badges or medals
  (\cite{KidMLB16})
  can help acknowledge the efforts. 
\end{itemize}

An extensive discussion on ``making replications mainstream''
is provided by Zwaan \etal (2018) \nocite{ZwaRELD18} and the 36 evoked comments.

\section{Conclusions}
\label{sec:concl}
The basic paradigm in science states that facts should only be recognized
as such if they can be and have been reproduced.
In many empirical science fields, this is not often practiced,
and it is difficult to judge which statements should be regarded as
reliable.
Even worse, when replication studies were undertaken, their results have
shown a disappointing rate of confirmation.
This insight has lead to the ``reproducibility crisis'' in large parts of
science.

An important trigger is
the urge to find new ``facts'' by a kind of raster screening.
The content of research is often not guided by interesting
relevant questions but by exploring many potential effects of minor
importance with the hope to find statistically significant ones in some
niche. The trap of selection bias or ``p-hacking'' snaps
(\cite{AmrVGM19}; \cite{WasRSL19}).

Awareness and concern about the problem have increased and lead to a flood
of editorials and articles---and even books---that have dealt with
diagnoses, interpretations, reviews,
and proposals for procedures and policies.
In this contribution, we have focussed on some basic issues:
\begin{itemize}
\item
  \textbf{Estimation, not testing.} \
Solid empirical research concerns important relations and aims at
estimation of relevant effects, intending not only to prove that they
are different from zero---which they are ``almost surely'' in any case.
An estimation problem asks for a relevance threshold, and the adequate
and straightforward way to present the result is by a confidence interval.
If a conclusion is needed, it should consist of checking whether
the confidence interval surpasses the threshold.
The small but important step from the misguided use of p values to
providing confidence intervals with their simple and direct relation to
the scientific problem of assessing an effect must finally be
consistently implemented.
\item
  \textbf{Between study variation.} \
  In any two studies, we should expect a variation between effects
  that is not restricted to the statistical variability of their estimates
  within each study as quantified by the formal standard error.
  This insight is best described by the random effects
  model of meta-analysis. 
  Consequently, large replication studies should not be
  expected to yield effect estimates that are compatible
  with the original in the sense of statistically
  insignificant difference, due both to this heterogeneity
  and to the selective reporting bias.

  The other important consequence is that
  the confidence interval obtained from a single study
  does not cover the true effect with the probability expressed by the
  nominal confidence level.
  It should be widened by a factor reflecting the between-study
  variance.
  A whole set of studies is needed to estimate this variance,
  or an informed value must be assumed. 
  Due to the selective reporting bias lurking in ``original studies,''
  sincere estimation is only achieved from pre-registered replications,
  thereby devaluating the quantitative results of the original study in
  favor of unbiasedness.

  This paradigm entails a fundamental change in planning replications.
  It is not really useful to conduct a single replication
  with a desired power, calculated on the basis of the assumption
  of equal true effects between original and replication. 
  Instead, a series of replications should be planned
  (possibly using a sequential design)
  and the sample sizes should be determined by power calculations
  respecting the heterogeneity, see \citeasnoun{PawSH20}.
\item
  \textbf{``Success'' of a replication.} \
  Statements about reproducibility should be careful in their use of
  terminology and differentiate possible outcomes.
  A binary answer is not helpful. For the replication of a significant or
  relevant effect in the original study, we suggest a classification
  with seven different outcomes. It is based on the two dimensions of
  relevance or significance of the effect in the replication and
  of the difference between the effects.
\end{itemize}

  Usually, a replication is designed as ``close'' as possible to the
  original, using, in the applicable sense,
  the same ``population'' and the same methods.
  This principle is meant to lead to a high probability of getting
  consistent results---which, as we just argued, will be ``successful''
  ($p<0.05$) only if the statistical power is kept low enough
  to avoid detection of the interstudy variability.
  For important scientific problems, however, an essential criterion
  is the generalizability of the result.
  Therefore, an adequate compromise between confirmation and extension
  of results is needed (see also conclusions in 
  \cite{MaxSLH15}).


In summary, then, a strategy is needed in order to obtain 
reliable scientific facts. 
An attempt to draft such a standard is the following.

\label{sec:strategy}
\begin{itemize}
\item
  If a claim is of basic interest for the field, multiple replications
  should be planned.
  A judgement is needed on the extent to which these replications should
  generalize the context and thus extend the domain of validity,
  and on the relevance threshold.
  These decisions may be a topic for professional societies.
\item
  For findings stemming from exploratory studies, 
  a first close (pre-registered) replication
  should be conducted, and depending on the result,
  more replications should follow:
  In the case of a confirmation ('Cnf' according to
  Section \ref{sec:ass.class}), generalizations can be studied,
  but in case of an attenuation or ambiguous result,
  more close replication is suggested.
  When replications are conducted without a pre-planned strategy,
  meta-analyses need to take the sequential aspects into account.
\item
  If a study serves to validate a theoretical proposition,
  it should be pre-registered in the first place.
\item
  In other cases, claims should be interpreted as working hypotheses
  (\cite{SorB89}  and others).
  Such exploratory results should still play an important role
  in science and, if done with enough care, get published as the potential
  source of replication or, more generally,
  as indications for generating theoretical hypotheses to be examined
  by pre-registered studies.
\end{itemize}
Such a strategy aims at structuring the process of knowledge generation.
They should avoid rules that restrict creativity, but rather help
distinguish the degrees of reliability of empirically based claims
and thereby save resources.

Our recommendations ask for substantial changes in the practice of
empirical research. 
We are convinced that the crisis of empirical science is even deeper than
recognized by the current discussion, and it is time to ask for the changes
needed to overcome it even though they sound overly challenging at present.
In the long run, solid establishment of scientific facts will
prove sustainable, whereas past and present practices dilute
the credibility of science and threaten to erode the support from society
it still enjoys.

\vspace{3cm}

\vspace{5mm} 
\textbf{Acknowledgement.}
Useful comments on an earlier version of the typescript have been provided
by Markus Kalisch.
Samuel Pawel gave very valualble hints to relevant literature and helped
strengthening arguments.

\pagebreak
\baselineskip14pt
\bibliography{./reprod.bib}

\pagebreak

\section{\Large
  Replicability: Terminology, Measuring Success, and Strategy\\ Supplmentary Material\\[7mm]
}
\author{Werner A. Stahel, \
Seminar for Statistics, ETH Zurich}

This document contains two supplementary sections to the paper on
Replicability: Terminology, Measuring Success, and Strategy.
\subsection{Bonnet's classification}

\citeasnoun{BonD20} classifies outcomes of a replication study based on the
confidence intervals for the effect $\IEff$\ and for the difference \IEDS,
too.
His classes are partly overlapping since their definitions are based on
either one of these intervals, which is reflected in the names containing
either ``directional'' (based on $\IEff$) or
``effect-size'' (based on their difference, \IEDS (!)), and one class
definition deviates from this (``Null effect-size nonreplication'' is based
on $\IEff$, see Table).

\newcommand{\CI}{\mathrm{CI}}
\newcommand{\aster}{$^*$\ }

{\small
\begin{tabular}{llp{16mm}p{27mm}p{58mm}}
Abbr&Eff.&Diff.&Name&Criteria\Hline{&&&&} 
repl.d & $\IEff>0$ & & Direc.\ repl. & 
  Reject both null hypotheses $H\sups o : \eff\sups o= 0$ and $H\sups o : \eff\sups r= 0$,
  same direction\\[5pt]
repl.es.S & & $\IEDS<h$ & Strong eff-s. repl.& 
  Accept $H\sups r : |\eff\sups o – \eff\sups r | < h$ \\[5pt]
non.d &$\IEff<0$ & &Direc.\ nonr. & 
  Reject both null hypotheses $H\sups o : \eff\sups o= 0$ and $H\sups o : \eff\sups r=0$,
  different directions \\[5pt] 
non.es.S &&$\IEDS>h$ or $\IEDS<-h$& Strong eff-s. nonr. & 
  The conf.int.\ for $\eff\sups o – \eff\sups r$ is completely outside
  the range of practical equivalence (AM-6)\\[5pt]
non.es.W & &$0\not\in\IEDS$& Weak eff-s. nonr. & 
  The conf.int.\ for $\eff\sups o – \eff\sups r$ excludes 0 (AM-5) \\[5pt]
non.d.Null\aster &$-h<\IEff<h$&& Null direc.\ nonr.\aster &
  Accept $H\sups r : |\eff\sups r | < h$ (AM-2)\\[5pt]
inc.d &$0\in\IEff     $ && Inconc.\ direc.& 
  The conf.int.\ for $\eff\sups r$ includes 0\\[5pt]
inc.es.S &&$h\in\IEDS$ or $-h\in\IEDS$&  Inconc.\ strong eff-s. & 
  The conf.int.\ for $\eff\sups o – \eff\sups r$ includes the values –h or h \\[5pt]
inc.es.W &&$0\in\IEDS$& Inconc.\ weak eff-s. & 
  The conf.int.\ for $ \eff\sups o – \eff\sups r$ includes 0 \\\hline
\end{tabular}
\aster: Has been named ``Null effect-size nonreplication'' (instead of
``directional'') in the paper
}

\citeasnoun{BonD20} further points to \citeasnoun{AndSM16}.
In that paper, different ``goals of replication'' are distinguished, and
partly agree with Bonnet's classes.
However, these classes cannot be ``goals'' since conducting a study for
just one of them would be rather unplausible. They are classes of results.

\pagebreak

\subsection{Data used as example in Section 5}
The data used for Figure 3 was extracted from the OSC15 publication
\cite{OpeSC15}:
The studies using a one sample or paired sample t test were selected and
two of them (115 and 122) dorpped because they contained some strange
features.
For the studies with negative original effect, the sign of test statistics
and estimated effects was inverted. The rows are sorted according to
decreasing original effect estimate.
This boils down to the data shown in Table \ref{tab:dataOsc}.

\begin{table}[h]
  \begin{tabular}{r|rrrrrrrrrr}
    Study & to & no & tr & nr & testType & EffSo & EffSr & eso & esr & eds \\ \hline  
 St.153 &  4.45 &   8 &  0.320 &   8 & single &  1.573 &  0.110 & 0.787 & 0.0566 & -0.7301 \\ 
 St.121 &  5.39 &  12 &  5.430 &  24 & paired &  3.250 &  2.220 & 0.778 & 0.5542 & -0.2238 \\ 
 St.7 & 10.18 & 100 &  0.496 &  15 & paired &  1.020 &  0.130 & 0.509 & 0.0640 & -0.4450 \\ 
 St.113 & 10.36 & 125 & 15.640 & 177 & single &  0.930 &  1.180 & 0.463 & 0.5878 &  0.1245 \\ 
 St.127 &  4.98 &  29 &  3.103 &  26 & paired & -0.940 & -0.620 & 0.462 & 0.3043 & -0.1579 \\ 
 St.146 &  3.20 &  15 &  1.900 &  12 & paired &  1.710 &  0.548 & 0.413 & 0.2742 & -0.1389 \\ 
 St.6 &  3.55 &  24 &  2.400 &  32 & paired &  0.724 &  0.424 & 0.362 & 0.2121 & -0.1502 \\ 
 St.33 &  3.77 &  40 &  2.080 &  40 & paired &  0.600 &  0.370 & 0.298 & 0.1644 & -0.1336 \\ 
 St.116 &  3.94 & 173 &  4.020 & 140 & paired &  0.470 &  0.340 & 0.150 & 0.1699 &  0.0201 \\ 
 St.15 &  1.93 &  95 &  3.955 & 242 & paired &  0.198 &  0.255 & 0.099 & 0.1271 &  0.0282 \\ 
\end{tabular}

\vspace{3mm}
\begin{tabular}{r|rrrrrrrrr}
  Study & ciwo & ciwr & ciwd & cidwo & cidwr & classo & classr & classeds & replclass \\ \hline  
 St.153 & 0.4180 & 0.4180 & 0.591 & 0.2956 & 0.2956 & Rlv & Amb & Rlv & Amb  \\ 
 St.121 & 0.2986 & 0.2111 & 0.366 & 0.2142 & 0.1515 & Rlv & Rlv & Amb & Cnf  \\ 
 St.7 & 0.1072 & 0.2769 & 0.297 & 0.0829 & 0.2140 & Rlv & Amb & Rlv & Amb  \\ 
 St.113 & 0.0883 & 0.0742 & 0.115 & 0.0626 & 0.0526 & Rlv & Rlv & Ctr & Enh  \\ 
 St.127 & 0.1912 & 0.2020 & 0.278 & 0.1353 & 0.1429 & Rlv & Rlv & Amb & Cnf  \\ 
 St.146 & 0.2841 & 0.3177 & 0.426 & 0.2012 & 0.2250 & Rlv & Amb & Amb & Amb  \\ 
 St.6 & 0.2082 & 0.1803 & 0.275 & 0.1476 & 0.1278 & Rlv & Sig & Amb & CnfW \\ 
 St.33 & 0.1599 & 0.1599 & 0.226 & 0.1131 & 0.1131 & Rlv & Sig & Amb & CnfW \\ 
 St.116 & 0.0752 & 0.0836 & 0.112 & 0.0532 & 0.0592 & Sig & Sig & Ngl & CnfW \\ 
 St.15 & 0.1011 & 0.0633 & 0.119 & 0.0733 & 0.0459 & Amb & Sig & Ngl & CnfW \\ 
\end{tabular}

\caption{\label{tab:dataOsc}
  Data underlying Figure 3 in the paper. The columns are
  \texttt{Study}: study number in OSC15;
\texttt{to, tr}: test statistics, original and replication;
\texttt{no, nr}: numbers of observations;
\texttt{testType}: type of t test: single sample or paired samples;
\texttt{EffSize.o, EffSize.r}: standardized effect as defined in OSC;
\texttt{eso, esr}: estimated effects according to this paper;
\texttt{eds}: estimated effect difference;
\texttt{ciwo, ciwr, ciwd}: width of confidence intervals for
  \texttt{eso, esr, eds};
\texttt{classo, classr, classeds}: class according to the classification of
  Section 3;
\texttt{replcalass}: class of the replication result according to Section
4.
}
\end{table}
\end{document}